# THE GREAT SEASON CLIMATIC OSCILLATION
# AND THE GLOBAL WARMING

*Ahmed Boucenna*

Laboratoire DAC, Département de Physique

Faculté des Sciences, Université Ferhat Abbas, 19 000 Sétif, Algeria

aboucenna@yahoo.com

## ABSTRACT

The present earth warming up is often explained by the atmosphere gas greenhouse effect. This explanation is in contradiction with the thermodynamics second law. The warming up by greenhouse effect is quite improbable. It is cloud reflection that gives to the earth's ground its 15 °C mean temperature. Since the reflection of the radiation by gases is negligible, the role of the atmosphere greenhouse gases in the earth warming up by earth radiation reflection loses its importance. We think that natural climatic oscillations contribute more to earth climatic disturbances. It is known that earth is subject to various climatic oscillations of relatively short periods such as the twenty-four hour and the one-year climatic oscillation periods. The other oscillation that we hypothesize to exist has a longer period (800 to 1000 years). The glacier melting and regeneration cycles lead to variations in the cold region ocean water density and thermal conductibility according to their salinity. These variations lead one to think about a macro climate oscillating between maximum hot and minimum cold temperatures. This oscillation is materialized by the passages of the planet through hot, mild, and cold eras, leading to the great season climatic oscillation phenomenon (GSCO). Thus, our planet lives four great seasons a great spring, a great summer, a great winter and a great autumn/fall, making a great year embracing our four small classical yearly seasons. This great season climatic oscillation is responsible for the slowing of the thermo haline circulation (THC) to the North Atlantic signaled by several authors. Culminating great summer heat weather maxima would take place around the years $N_s$ = 2000 ± (800 to 1000)k and culminating great winter cold weather maxima would be expected to occur around the years $N_w$ = 1600 ± (800 to 1000)k, where k is an integer number. The probabilistic character of the parameters that are at the origin of this climatic oscillation makes the long-term prediction less precise but the deterministic tendencies and the resonance phenomena give precious information on our planet climatic future. The present warming up is well interpreted in the frame of the great season climatic oscillation theory. The macro climate of the Maghreb and Europe countries is strongly influenced by this great season climatic oscillation. The north inhabitants spend difficult periods during the great winter coldest period, while those living in arid regions live difficult moments during the great summer hottest period, with drought and lack of water for people and animals. History, sociology, peoples and migration flux are also influenced by these great season climatic oscillations.

## 1. INTRODUCTION

The earth's ground mean temperature is estimated at 15 °C. The global warming up is appreciated by the evaluation of the fluctuation of the earth's ground mean temperature around this 15 °C mean value. A previous calculation [1] based on the black body radiation theory gives an earth's ground mean temperature of the order of -18 °C which is much lower than 15 °C. The important gap between the mean values calculated and estimated requires an explanation. The greenhouse effect assigned to the atmosphere greenhouse gases (GHG) has been put forward to explain this gap [2-9]. This earth warming up explanation is in contradiction with the thermodynamics second law which stipulates that a colder body cannot warm a hotter one. The warming up by greenhouse effect is quite improbable. In this work, we will recalculate the earth's ground mean temperature to show the importance of the role of the earth IR radiation reflection by the atmosphere clouds (solid and liquid). It is the cloud reflection and not the greenhouse effect due to the atmosphere greenhouse gases that gives to the earth's ground its mean temperature. The 15 °C mean value is obtained by choosing adequate values for the sun and earth radiation reflection coefficients. Since the reflection of the radiation by gases is negligible, the role of the atmosphere greenhouse gases in the earth warming up by earth radiation reflection loses its importance. The natural climatic oscillations contribute more to the earth climatic disturbances. The present warming up is rather explained by a longer period climatic oscillation that we believe to exist. The glacier melting and regeneration cycles lead to variations in the cold region ocean water density and thermal conductibility according to their salinity. These variations lead one to think about a macro climate oscillating between maximum hot and minimum cold temperatures. This oscillation is materialized by the passages of the planet through hot, mild, and cold eras, leading to the great season climatic oscillation phenomenon (GSCO) [10].

## 2. EARTH'S GROUND MEAN TEMPERATURE

### 2.1. Power coming from the sun

The power radiated by a T temperature surface is given by the Stefan-Boltzmann law:

$$P = \sigma T^4 \tag{1}$$

where $\sigma$ is the Stefan constant given by:

$$\sigma = \frac{2\pi^5 k^4}{15 c^2 h^3} = 5.6704 \, 10^{-8} \, \frac{W}{m^2 K^4} \tag{2}$$

$c$ is the light speed, $h$ and $k$ are respectively, the Planck and Boltzmann constants. Thus:

$$P \approx 5.67 \left(\frac{T}{100}\right)^4 \, \frac{W}{m^2 K^4} \tag{3}$$

For $T_{sun}$ = 5780 K, the power S received by the earth's atmosphere from the sun at the equator is:

$$S = \sigma T_{sun}^4 \, \frac{A_{sun}}{A_{earth\_orbit}} = \sigma T_{sun}^4 \, \frac{R_{sun}^2}{R_{earth\_orbit}^2} \tag{4}$$

where $R_{sun}$ and $R_{earth\_orbit}$ are respectively the sun and the earth orbit radii and , $A_{sun}$ and $A_{earth\_orbit}$ are the sun and the earth orbit areas. Therefore, 1368 J of solar electromagnetic radiation arrive every second perpendicularly on every $m^2$ of the high earth's atmosphere. These radiations are essentially composed of visible light 44.8 %, infrared radiation 45.2% and ultraviolet radiation 10%. A first part of this energy is reflected toward the space, a second part is absorbed by the atmosphere, and another arrives to the earth surface to cover all earth and earth inhabitant needs.

### 2.2 Earth ground mean temperature without reflection

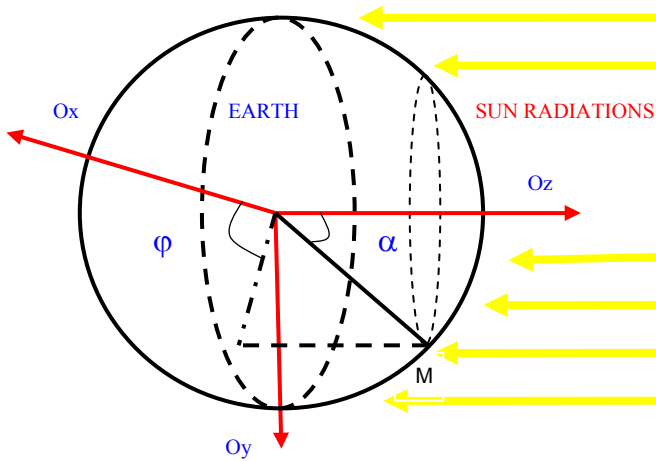

Fig.1. Exposed Earth face. The $\alpha$ latitude is the angle between the Oz axis (passing by the earth center and crossing the equator) and the vector locating the solar radiation impact point (on the earth's ground) M.

At any $\alpha$ latitude point of the exposed earth face of the atmosphere, the power coming from the sun (Fig. 1) is:

$$P = S \cos \alpha \tag{5}$$

If the earth behaves like a perfect black body, the earth's ground absorbs all this power. The earth is in equilibrium with its environment, it must radiate the power that it absorbs. The earth's ground temperature must satisfy the relation:

$$\sigma T_{earth\_ground}^4 = S \cos \alpha \tag{6}$$

From where the earth's ground temperature will be given by:

$$T_{earth\_ground}^4 = \frac{S}{\sigma} \cos \alpha \qquad (7)$$

and:

$$T_{earth\_ground} = \sqrt[4]{\frac{S}{\sigma} \cos \alpha} \qquad (8)$$

Table 1. Earth's ground temperature distribution versus exposed earth face $\alpha$ latitude according to relation (8).

| Region | Latitude $\alpha$ | $T_{earth\_ground}$ (K) | $T_{earth\_ground}$ (°C) |
|---|---|---|---|
| Equator | 0° | 394.09 | 121.09 |
| | 10° | 392.50 | 119.50 |
| | 20° | 387.93 | 114.93 |
| | 30° | 380.16 | 107.16 |
| MAGHREB | 40° | 368.68 | 95.68 |
| E | 45° | 361.36 | 88.36 |
| U | 50° | 352.89 | 79.89 |
| R | 60] | 331.38 | 58.38 |
| O | 70° | 301.37 | 28.37 |
| P | | | |
| E | 80° | 254.52 | -18.47 |
| Pole | 90° | 0.00 | -273.00 |

The mean power, $P_{mean}$, radiated on the whole earth's ground is:

$$P_{mean} = \frac{1}{A} \int_{exposed\_face\_area} S \cos \alpha \, dA \qquad (9)$$

where $A = 4\pi R^2$ is the earth area, thus :

$$P_{mean} = \frac{SR^2}{4\pi R^2} \int_0^{2\pi} \int_0^{\frac{\pi}{2}} \cos \alpha \sin \alpha \, d\alpha \, d\varphi = \frac{S}{2} \int_0^{\frac{\pi}{2}} \cos \alpha \sin \alpha \, d\alpha \qquad (10)$$

Let us calculate: $\int_0^{\frac{\pi}{2}} \cos \alpha \sin \alpha \, d\alpha$. Putting: $u = \cos \alpha$, then: $du = -\sin \alpha$, and:

$$\int_0^{\frac{\pi}{2}} \cos \alpha \sin \alpha \, d\alpha = -\int_1^0 u \, du = \int_0^1 u \, du = \left[ \frac{1}{2} u^2 \right]_0^1 = \frac{1}{2} \qquad (11)$$

Then:

$$P_{mean} = \frac{S}{4} \qquad (12)$$

If $T_{mean}$ is the earth's ground mean temperature, one has:

$$\sigma T_{mean}^4 = P_{mean} \qquad (13)$$

and:

$$T_{mean} = \sqrt[4]{\frac{1}{\sigma}\frac{S}{4}}$$ (14)

which gives: $T_{mean} = 278.66 \text{ K} = 5.66\,°C$. This is the earth's ground mean temperature if there were no reflection by the earth.

## 2.3 Earth's ground mean temperature with reflection

The earth and its atmosphere present a reflection coefficient R estimated at $R \approx 0.30$. Therefore only 70 % of the power $P_{mean}$, coming from the sun reaches the earth ground exposed face:

$$P_0 = (1-R)\,P_{mean} = (1-R)\,\frac{S}{4}$$ (15)

If this is the absorbed power that gives the earth's ground its mean temperature, $T_{mean}$, that satisfies:

$$(1-R)\,\frac{S}{4} = \sigma\,T_{mean}^4$$ (16)

Then:

$$T_{mean}^4 = (1-R)\,\frac{1}{\sigma}\frac{S}{4}$$ (17)

and:

$$T_{mean} = \sqrt[4]{(1-R)\,\frac{1}{\sigma}\frac{S}{4}}$$ (18)

Thus : $T_{mean} = 254.89 \text{ K} = -18.10\,°C$. This result agrees with previous calculation [1] based on the black body radiation theory which gives an earth's ground mean temperature of the order of -18 °C. In fact the relation (16) does not give the whole mean absorbed power by the earth. The earth radiates toward the space, as IR radiations, the power $P_0$ that it absorbed. This infrared radiation $P_0$ emitted by the earth, tries to cross the interface (Earth+atmosphere)-space characterized by a reflection coefficient r. Only a part $(1-r)\,P_0$ crosses this interface and leaves the earth, whereas the power:

$$P_1 = r\,P_0$$ (19)

is reflected to the earth which absorbs and reemits this power to the space as IR radiations. The interface will then reflect to the earth the power:

$$P_2 = r\,P_1 = r^2\,P_0$$

and so on. At the end of the $n^{th}$ reflection, the power reflected by the interface is:

$$P_n = r\,P_{n-1} = r^n\,P_0$$

Therefore, the power effectively received by earth is:

$$P = P_0 + P_1 + P_2 + P_3 + \ldots + P_n = P_0 + r^1\,P_0 + r^2\,P_0 + \ldots + r^n\,P_0$$ (20)

P is the sum of n first terms of a geometric progression, where the first term is $P_0$ and the reason is r. Therefore:

$$P = \frac{r^{n-1}-1}{r-1}\,P_0$$ (21)

For n $\rightarrow \infty$, one has :

$$P = \frac{1}{1-r} P_0 \qquad (23)$$

Therefore, the power effectively absorbed by the earth's ground and reemitted as IR radiations is:

$$P = \frac{1-R}{1-r} \frac{S_0}{4} \qquad (24)$$

This is the absorbed power that gives the earth's ground its mean temperature, $T_{mean}$, that satisfies:

$$\frac{1-R}{1-r} \frac{S}{4} = \sigma T_{mean}^4 \qquad (25)$$

Therefore:

$$T_{mean}^4 = \frac{1}{4} \frac{1-R}{1-r} \frac{S}{\sigma} \qquad (26)$$

and:

$$T_{mean} = \sqrt[4]{\frac{1}{\sigma} \frac{1-R}{1-r} \frac{S}{4}} \qquad (27)$$

For R = 0.31 and r = 0.40 one has:

$$T_{mean} = 288.57 \text{ K} = 15.57 \text{ °C}$$

Relation (27) gives the earth's ground mean temperature, if the earth, with its atmosphere, its oceans, its inhabitants, its temperature and pressure gradients, … did not keep any power. In fact, a fraction $S_u$ of this power is consumed by the earth. The power serving to the ocean water evaporation is estimated at 78 W/m$^2$ s and the one serving to heat the air is 24 W/m$^2$ s. If one takes into account this power loss that the earth keeps, the relation (26) becomes:

$$T_{mean}^4 = \frac{1}{\sigma} \left[ \frac{1-R}{1-r} \frac{S}{4} - S_u \right] \qquad (28)$$

and:

$$T_{mean} = \sqrt[4]{\frac{1}{\sigma} \left[ \frac{1-R}{1-r} \frac{S}{4} - S_u \right]} \qquad (29)$$

For $S_u$ = 102 W/m$^2$, R = 0.31 and r = 0.521 one has:

$$T_{mean} = 288.08 \text{ K} = 15.08 \text{ °C}$$

The $T_{mean} = 288$ K = 15 °C earth's ground mean temperature has been obtained while considering an earth IR reflection coefficient r equal to 0.521. This reflection coefficient cannot come from the atmosphere gases ($O_2$, $N_2$, A, $H_2O$ (steam), $CO_2$, Kr, H, $N_2O$, Xe, $O_3$ (ozone), $CH_4$). Gases have a very weak reflection coefficient because of their refractive index value n ~ 1. The reflection coefficient r for a radiation at the interface between two media of refractive indexes $n_1$ and $n_2$ is given by the relation:

$$r = \left( \frac{n_1 - n_2}{n_1 + n_2} \right)^2 \qquad (30)$$

The reflection at the interfaces gas-vacuum and gas-gas is negligible: $n_2 - n_1 \approx 0$ since $n_2 \approx n_1 \approx 1$. Gases do not have a reflection coefficient. They do not reflect light since the molecule sizes are too small compared to the radiation wave length. These are the sprays, that are microscopic particles hanging on the air, and the clouds composed of water droplets (liquid) or water solid (ice) that can reflect and diffuse the radiations coming from the earth or from the sun. At the interface between the earth's ground and the atmosphere, the IR radiation emitted by the earth is reflected with a reflection coefficient $r_e$ estimated at 0.31. The total reflection coefficient r for earth's IR radiations that results from the contribution of the interface earth-atmosphere with reflection coefficient $r_e$ and of the interfaces: vacuum-gas, gas-cloud and all others sprays with reflection coefficient $r_c$, is:

$$r \approx r_e + r_c \tag{31}$$

Assuming that the reflection coefficient $r_c$ is practically equal to the cloud reflection coefficient, one can deduce an approximate value of $r_c$:

$$r_c \approx r - r_e \approx 0.21 \tag{32}$$

The reflection coefficient $r_c$ is linked to the atmosphere composition (clouds, smoke, dust …). The clouds having a strong reflection coefficient reflect a good part of earth's IR radiations toward the earth's ground and that is what keeps the night earth's ground temperatures moderate in the presence of clouds. On the other hand when the sky is cleared, the reflection coefficient is weak and the reflection of earth's IR radiations is less important giving colder nights. The effect of the reflection coefficient of the clouds that reflects the heat during the night is evident. If there were no other factors that influence the earth climate, the exposed earth face temperature varies from one point to another versus the $\alpha$ latitude as given by relation (8) and shown in table 1. Therefore, the power radiated by the sun is not uniformly distributed on the earth's ground. The poles temperatures are lower than the equator ones. Other mechanisms are in charge for a more interesting distribution.

## 3. THE DETERMINIST CLIMATIC CYCLES
### 3.1. The earth rotation

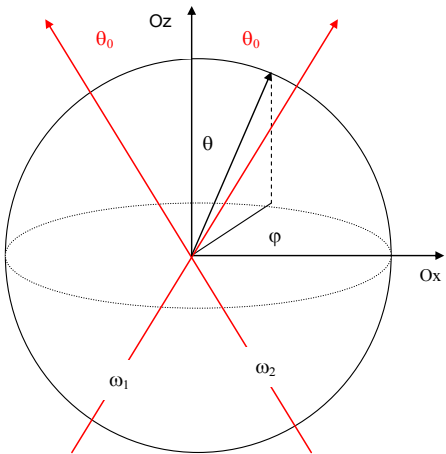

Fig. 2 Earth rotation and incline. The spherical coordinates $\theta$ and $\varphi$ locate an impact earth's ground point of the solar radiation, with respect to the Ox and Oy axes (passing by the earth center and the equator) and the Oz axis (passing by the North and the South poles). The limits of the earth Incline are indicated by the axis $\omega_1$ and $\omega_2$.

One introduces geometrical and dynamical factors to take into account the earth rotation and incline with angular frequencies $\omega_d$ and $\omega_y$ corresponding to the night-day and year cycles (Fig. 2). For the rotation (night-day cycle) one has:

$$\sigma T_{earth\_ground}^4 = \begin{cases} \dfrac{1-R}{1-r} \, S \sin\theta \cos(\varphi - \omega_d t) & \text{if} \quad -\pi/2 \leq \varphi - \omega_j \leq \pi/2 \\ 0 & \text{if} \quad \pi/2 \leq \varphi - \omega_j \leq 3\pi/2 \end{cases} \tag{33}$$

### 3.2. The North-South axis incline

When including the obliquely rotating earth (year cycle), the formula is more complicate:

$$\sigma T_{earth\_ground}^4 = \begin{cases} \dfrac{1-R}{1-r}\, S\, L(\theta_0, \theta, \varphi, \omega_y, \omega_d, t) & \text{exposed face} \\ 0 & \text{not exposed face} \end{cases} \quad (34)$$

where [1]:

$$L(\theta_0, \theta, \varphi, \omega_y, \omega_d, t) = [\sin(\omega_y t)\cos(\omega_d t) + \cos(\omega_y t)\sin(\omega_d t)\cos(\theta_0)]\sin\theta\cos\varphi$$
$$+ [-\sin(\omega_y t)\sin(\omega_d t) + \cos(\omega_y t)\cos(\omega_d t)\cos(\theta_0)]\sin\theta\sin\varphi \quad (35)$$
$$- [\cos(\omega_y t)\sin\theta_0]\cos\theta$$

## 4. THE PROBABILISTIC CLIMATIC CYCLES

The earth climate is mainly modulated by determinist oscillations linked to the earth rotation and the North-South axis incline, leading to a given distribution of the heat received from the sun. This distribution remains very insufficient for a harmonious life development on the planet. The intervention of other factors is necessary. The atmosphere, the oceans, the seas, the mountains … are here to make the earth an ideal place for the man and its environment. The atmosphere influences, in a probabilistic way, in the making of the climate since the phenomena that intervene are quite random. The oceanic streams play an important role in the making of the earth climate:
- at short-term, by contributing in a meaningful way to moderate the climate of the cold regions as Europe,
- and at long-term, by the great season climatic oscillation phenomenon (GSCO) whose factors are probabilistic but whose cycle is quite determinist.

## 4.1. The Great Season Climatic Oscillation (GSCO)

The oceanic streams, named "thermo haline circulation" or THC, are the gigantic " heat marine rivers" that browse the oceans. The glacier melting and regeneration cycles lead to variations in the cold region ocean water density and thermal conductibility according to their salinity. These variations lead one to think about a macro climate oscillating between maximum hot and minimum cold temperatures. The hot streams formed of hot and low density water go on surface, from hot to cold regions to warm them. In cold regions, water becomes colder with high density due to high cold region salinity. Colder water with higher density dive deep in oceans giving cold and deep high density streams, which return to hot regions to close the circuit. A depth d separate surface hot streams going from hot to cold regions and deep cold streams moving from cold to hot regions. At a given moment, the cold streams are to the maximal limit. While surface hot streams arrive in cold region, the glacier melting increases and cold water salinity and density decrease. Therefore, low density cold water dives less deeply in oceans. Lower density cold streams then progressively come closer to the surface to interfere with surface hot streams. Hot streams effects are then progressively compensated by cold streams. Hot streams are progressively decreased, slowed down or even stopped. Glaciers regeneration is favoured again. Once more, a deeper cold stream progressively moves away from surface hot streams. Cold streams influence on hot ones decreases. Hot streams arrive again to cold regions provoking again glaciers melting and so forth. The depth $d$ separating hot and cold streams is then not constant. It oscillates between two limiting values $d_1$ and $d_2$ according to cold region salinity $S$ that oscillates also between two limiting values $S_1$ and $S_2$ corresponding to optimal melting and glaciers regeneration.

The measures concerning the central Labrador Sea water salinity versus the depth and the time, reported in reference [11] show that the salinity of the deepest part of the water column has decreased for some decades. This reduction entails the slowing of the deep water formation that is the downward part of the thermo haline circulation.

In reference [12] it is shown that the thermo haline circulation is slowing. A tendency to the decrease, estimated between 2 % and 4 % per year, since the middle of the 1990 – years, of the water debit that passes through the Faroe strait (Faroe Shetland Channel) under the 0.3 °C isotherm, is noted. The deep circulation coming from the sea of Norway (where one part of the deep water is formed), that dives in the deep Atlantic, seems to decrease now. These observations agree with the great season climatic oscillation theory predictions [10].

This oscillation (melting and regeneration of the glaciers) materialized by the passages of the planet through hot, mild, and cold eras, entails the great seasons climatic oscillation phenomenon. Thus, our planet lives four great seasons a great spring, a great summer, a great winter and a great autumn/fall, making a great year embracing our four small classical yearly seasons. The great season climatic oscillation theory is based on the dependence between the depth of the cold oceanic streams and the salinity (therefore the density) of the cold ocean region waters. Culminating great fall is reached when the cold streams are in the closest position of the hot streams. The influence of the cold streams on the hot streams is maximal. The hot streams are slowed down to a minimum. The regeneration of the glaciers is optimal and announces the great winter arrival. Culminating great spring corresponds to the farthest position of cold from the hot streams. The hot stream intensity is maximal. The melting of the glaciers is optimal announcing the approach of the great summer. Culminating great winter and great summer correspond to a middle position.

### 4.2. Great season climatic oscillation period
The thermo haline circulation period is roughly estimated at 1000 years. One could assimilate the great season climatic oscillation period to this one. With the lack of all necessary data to exactly determine this climatic oscillation period generating great seasons, assuming the variation to be sinusoidal, and based on the thermo haline circulation period and on historical observations, we estimate this period to be equal to eight to ten centuries: $T = 800 \text{ to } 1000 \text{ years}$.

The length of a great season would be equal to two or two and half centuries (200 to 250 years). The process of the great season climatic oscillations is very slow. It constitutes the basis of the mechanism of the changes at very long term of the earth climate. One may not perceive the existence of the great season climatic oscillation since they are generated by a very slow physical phenomenon compared to that he could observe along his life. Their effects would appear to him as being temporary climatic changes whereas they are actually periodic phenomena that repeat themselves during sufficiently well determined periods.

### 4.3. The probabilistic factor Impact
The phenomena intervening in the great season climatic oscillation cycle have a random character. The intensity, the temperature and the depth of the streams, the melting and the regeneration of the glaciers, the density and the quantity of waters that participate do not obey determinist laws since some probabilistic phenomena (statistics) intervene. The GSCO cycle, although determinist has a quite variable period.

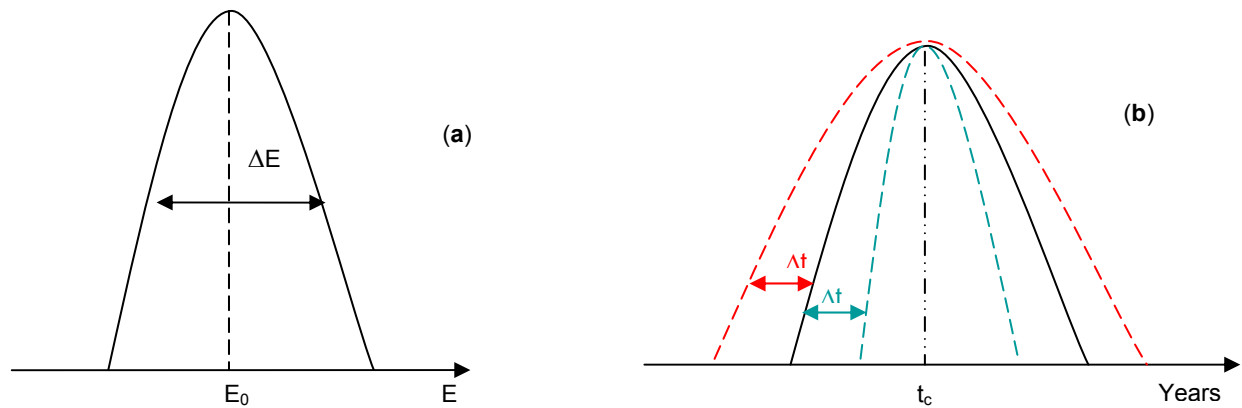

Fig. 3 Gamma radiation peak Widening due to probabilistic factors (**a**). Widening and shortening of a great climatic season period due to probabilistic factors (**b**)

Like all phenomena bound to probabilistic parameters, the great season climatic oscillation is exposed to a temporal widening or shortening. A peak obtained, for example, by the detection of a given energy gamma radiation has a width, due to the fact that the reactions of the radiation in detector are probabilistic. The peak has a width $\Delta E$ around the value $E_0$ corresponding to the energy of gamma radiation (Fig. 3 **a**). In the same way, because of the probabilistic factors the length of a great season climatic oscillation could be enlarged or shortened in time (Fig. 3 **b**). This implies a quite variable period for the great season climatic oscillation. Because of the probabilistic factors, a displacement is probably introduced after every cycle. The periods of the great season climatic oscillation, and therefore the great year efficient lengths can vary from a great year to another.

### 4.4. The resonance phenomena
If an oscillator is excited by a signal with a frequency equal to the eigen frequency of this oscillator, the oscillator vibration amplitude is amplified. It is the resonance phenomenon. If a great season settles, a great summer for example, the eigen season (or the eigen frequency) of the planet is the summer. The small seasons constitute excitation. Then, if an excitation has the same frequency (the same season) one has an amplification of the season effects. Therefore during a great summer the summer effects are amplified. In the same way during the great fall the fall effects are amplified, etc.

## 5. RESULTS AND DICUSSION
### 5.1. The climatic determinist tendencies

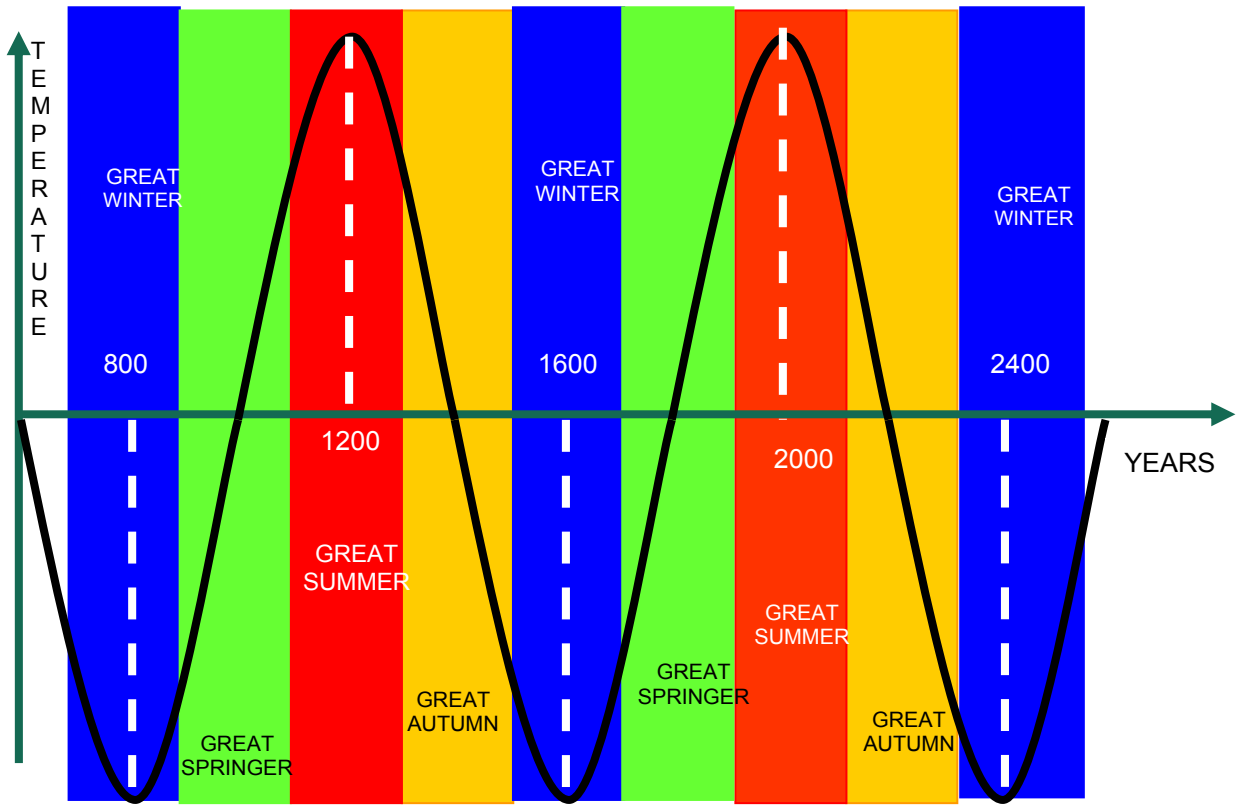

Fig. 4. Sinusoidal great season climatic oscillation

The great season climatic oscillations phenomenon is a natural phenomenon, specific to our planet. It is due to the presence of salts in the ocean waters. This climatic oscillation is determinist by its cycle (melting – regeneration - melting of glaciers) that maintains itself. The melting and the regeneration of the glaciers control the cold region water density variation (i.e. salt concentration of the cold region waters). This determines the cold current depth that controls the hot current intensity that favors melting or regeneration of glaciers. The great season climatic oscillation period is the fundamental parameter for the predictions of the world macro climatic changes at long-term. Culminating great winter cold weather maxima would be expected to occur around the years:

$$N_W = 1600 \pm (800 \text{ to } 1000) \, k \tag{36}$$

where k is an integer number. $N_W$ corresponds then to:

$$-2400, -1600, -800, 1, 800, 1600, 2400, 3200 \dots$$

The last great winter coincides with the last mini ice age that Europe experienced and which contemporary writers have testified. The North inhabitants had difficult moments during the coldest part of the great winter. During this great season the glaciers (of the north) extend to South to cover a large part of Europe. In the Maghreb the snows fall abundantly and cover the mountains during all year. The passage by the great winter allows the earth to regenerate its resources in fresh water. The abundant snows that fall permit the regeneration of the water underground. So, poor water regions, such as the Maghreb and the Middle East, regenerate their reserves. Dry rivers, of the last Great Summer, start flowing again. Forest regenerations in terms of trees population and others are observed. Culminating great summer heat weather maxima would take place around the years:

$$N_s = 2\,000 \pm (800 \text{ to } 1000) \, k \tag{37}$$

Corresponding to :

$$-3600, -2800, -2000, -1200, -400, 400, 1200, 2000, 2800, 3600, \dots$$

As shown in references [13-14], concerning northern hemisphere mean temperature during the past 11 000 years, historic data show such alternating warming up and cooling periods which affected the earth since thousands of years.

**5.2 The global Warming**

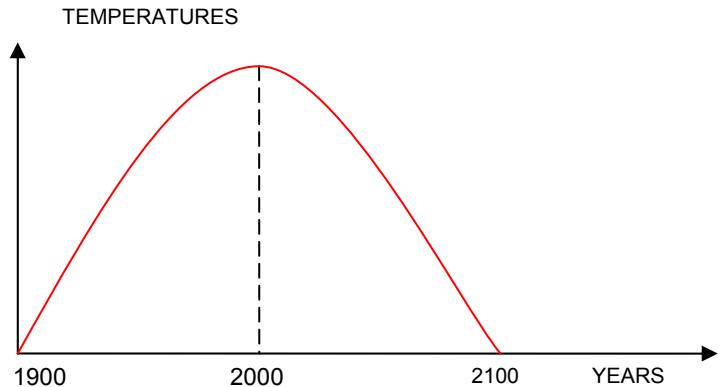

Fig. 5. Earth's ground temperature variation predicted by the great season climatic oscillation versus years

For the time being, a great summer is being experienced. During the great summer, the glaciers shrink northbound to cover only the north of Europe. In the Maghreb, snow falls only very little and covers the mountains only some days in the year. This explains the present warming up. The increase of the temperature observed since 1880 [2] is well interpreted in the frame of the great season climatic oscillation theory (fig. 5). The arid region inhabitants live difficult moments during the hottest period of the great summer, with drought and lack of water for people and animals. Some species can even disappear. History, sociology, people migration flux are also influenced by these great season climatic oscillations.

**5.3. Relocation of the hot and cold Centers**
The relocation of the well-known centers to be HOT or COLD of the planet, during a large part of the great summer and the great fall, provokes particular and unusual climatic disruptions.

**6. CONCLUSION**
In this work, we recalculated the earth's ground mean temperature to show the importance of the role of the earth IR radiation reflection by atmosphere cloud (solid and liquid). This cloud reflection, and not the greenhouse effect due to atmosphere greenhouse gases, gives to the earth's ground its mean temperature. The 15 °C earth's ground temperature mean value is obtained by choosing adequate values for the sun and earth radiation reflection coefficients. Since the reflection of the radiation by gases is negligible, the role of the atmosphere greenhouse gases in the earth warming up by earth radiation reflection loses its importance.

The present warming up is explained by a great season climatic oscillation. The cycle of glacier melting and regeneration entails variations of the density and the thermal conductibility of the cold region ocean waters according to their salinity leading to the oscillation of the macro climate between two extreme positions: a maximum of hot and a minimum of cold temperatures. This oscillation results in the passage of the planet per hot, mild and cold eras and leading to the great season Climatic Oscillation phenomenon.

Thus, our planet lives four great seasons a great fall, a great winter, great spring, and a great summer making a great year embracing our four small classical yearly seasons. The great season climatic oscillation period is estimated to be equal 800 to 1000 years. This climatic oscillation is responsible of the North Atlantic thermohaline circulation slowing signaled by several authors [15]. The great season Climatic Oscillation phenomenon is linked to some random factors but is quite deterministic; it intervenes in the making of the final earth macro climate. The probabilistic character of the parameters that are at the origin of this climatic oscillation makes the long-term prediction less precise but the deterministic tendencies and the resonance phenomena give precious information on our planet climatic future.

**Acknowledgement**
Thanks are due to Professor A. Layadi for his help.